\documentclass[twocolumn,apj,dvipdfmx]{emulateapj}
\usepackage{color}
\usepackage{ulem,cancel,amsmath}
\def\mbf#1{\mbox{\boldmath ${#1}$}}

\shorttitle{Dust Dynamics in Protoplanetary Disk Winds}
\shortauthors{Miyake, Suzuki \& Inutsuka}

\begin{document}

\title{Dust Dynamics in Protoplanetary Disk Winds Driven by
Magneto-Rotational Turbulence: A Mechanism for Floating Dust Grains with
Characteristic Sizes}

\author{Tomoya Miyake, Takeru K. Suzuki, \& Shu-ichiro Inutsuka}
\email{miyake.tomoya@e.mbox.nagoya-u.ac.jp; stakeru@nagoya-u.jp}
\altaffiltext{}{Department of Physics, Nagoya University, Nagoya, Aichi 
  464-8602, Japan}

\begin{abstract}
  We investigate the dynamics of dust grains with
  various sizes in protoplanetary disk winds driven by magnetorotational turbulence, by 
  simulating the time evolution of
  the dust grain distribution in the vertical direction. 
  Small dust grains, which are well coupled to the gas, 
  are dragged upward with the up-flowing gas, 
  while large grains
  remain near the midplane of a disk.
 Intermediate--size grains float near the sonic point
 of the disk wind located at several scale heights
 from the midplane, where
 the grains are
 loosely coupled to the background gas.
  For the minimum mass solar nebula at 1 AU, dust grains with size of
  25 -- 45 $\mu m$ float around 4 scale heights
  from the midplane. 
  Considering the dependence on the distance from the central star, smaller-size 
  grains remain only in an outer region of the disk, while larger-size grains 
  are distributed in a broader region. 
  We also discuss the implication of our result to the observation of dusty material
  around young stellar objects.
\end{abstract}

\section{Introduction}

Protoplanetary disks are believed to be the birth place of planets
\citep[e.g.,][]{saf72,hay76,hay85,cam78}, and have been 
extensively studied from both theoretical and observational sides
\citep[e.g.,][and references therein]{arm11}.
One of the major unknowns in the evolution of protoplanetary disks 
is the dispersal mechanism at the final stage.
From near infrared observations the typical dissipation time of protoplanetary disks is inferred to be $10^6-10^7$ yr \citep[][]{hai01,her08,yas09,tak14,tak15}. 
The photoevaporating wind by UV\citep[e.g.,][]{hol94,ale06}
and X-ray \citep[e.g.][]{erc08} emissions from a central star has been 
extensively investigated.
The disk wind driven by magneto-turbulence,
which is the magneto-rotational instability
\citep[MRI hereafter;][]{vel59,cha61,bal91} in a disk,
is also proposed as a cooperating process or a powerful alternative
in dispersing the gaseous component \citep{SI09}.

In order to
promote the mass accretion by the transport of the angular momentum 
in a protoplanetary disk \citep[]{lyn74}, the gas has to be in a turbulent 
state at least in some portions of the disk to keep the effective turbulent 
viscosity since the microscopic viscosity is negligible. 
Among various mechanisms, MRI is supposed to be a promising 
candidate to excite such turbulence. 
\citet{SI09} showed that the MRI-
driven turbulence inevitably induces
vertical outflows by the Poynting flux of the magnetohydrodynamical turbulence 
\citep[see also][]{ms00}, and such disk winds potentially contribute 
to the dispersal of the gas component of protoplanetary disks \citep{SMI10}, 
whereas this processes has been extended \citep{ogl12,bs13a,les13} and the 
quantitative mass flux carried by the disk wind is still under debate 
\citep{fro13}.

A spectro-astrometric survey of protoplanetary disks by
\citet{pon11} showed non-Keplerian motions that are consistent with wide-angle
disk winds, in addition to the component of Keplerian rotation. 
This might be an indirect
evidence of such MRI turbulence-driven vertical outflows.
\citet{nat14} also reported the observations of
slow gaseous outflows with a few km s$^{-1}$ from 
protoplanetary disks around T-Tauri stars.
\citet{ell14} observed time-variable optical fadings and infrared brightenings, 
which is supposed to reflect dust grains lifted up high above the disk 
midplane by a slow wind. 
Measurement of hydrogen atoms toward young stars show the
absorption
of the Lyman-$\alpha$ band in excess of that is inferred from the circumstellar 
extinction
\citep[]{McJ14}, which may be also caused by dust grains 
stirred up to the upper layers.
Near-infrared imaging of RY Tau suggested that the disk could
have an optically thin and geometrically thick layer above the disk surface
\citep[]{tkm13},
which may reflect dust grains floating at the upper atmosphere.

The dynamics of dust grains in the MRI turbulence-driven disk winds
should be potentially interesting in the context of planetesimal formation
but have not been studied yet.
In this paper,
we investigate the time evolution of the vertical 
distribution of dust grains with various sizes in the disk wind by simple 
one-dimensional (1D) simulations.

\section{Model}

We compute
the time evolution of
the vertical profile of dust density 
(Section \ref{sec:dust}) under a fixed background gas component 
(Section \ref{sec:gas}) in a (1D) vertical box from 
$z=-10 H_0$ to $z=10 H_0$, 
where $H_0=\sqrt{\mathstrut 2}c_{\rm s}/\Omega_{\rm k}$ is the pressure scale 
height derived from the sound speed, $c_{\rm s}$, and the Keplerian rotation 
frequency, $\Omega_{\rm K}$.
In order to avoid unphysical effects at the boundaries, we set up 
an extra $\pm 5H_0$ region outside the top and bottom boundary, and at 
$z=\pm 15 H_0$ the outflow boundary condition is prescribed.
As the initial condition, all the dust 
grains are sedimented at the midplane, $z=0$, and the total mass 
of the dust in the computational domain is 1/100 of the total mass of the gas.
 We use the normalization of $\Omega_{\rm k}=1$ and 
 $H_0=1$
 , which gives 
$c_{\rm s}^2=1/2$.

\subsection{Dust Component}
\label{sec:dust}
Dust grains are treated as
``pressure-less fluid''
in a fixed background gas 
distribution that is modeled in Section \ref{sec:gas}, namely 
drag force is exerted on the dust fluid from the background gas component, 
but the back reaction from the dust to the gas is not taken into account.
In this work, we consider dust grains of which size is smaller than a mean 
free path of the gas. 
In this case, the gas drag force is in the Epstein's regime \citep{ahn76,sek83,paa07}. 
For example, in the minimum mass solar nebula 
\citep[MMSN hereafter; ][]{hay81}, the mean free path is $>$ 1 cm for 
distance $> 1$ AU from the central star \citep[]{nak86};
the dust grains we consider here is smaller than this value.
Then, the drag force is written as \citep{sch63}
\begin{eqnarray}
 \mbf{F}_{\rm D}&=&
  -\pi\rho_{\rm g}a^2|\Delta\mbf{v}|\Delta\mbf{v} \nonumber \\
 & &\times{\scriptstyle  \left[\left(1+\frac{1}{{\cal M}^2}-\frac{1}{4{\cal
				M}^2}\right){\rm erf}({\cal M}) +\left(\frac{1}{{\cal M}}+\frac{1}{2{\cal M}^3}\right)\frac{{\rm
				e}^{-{\cal M}^2}}{\sqrt{\pi}}
				    \right]},
 \label{eq;epsori}
\end{eqnarray}
where $\rho_{\rm g}$ is the gas density, $a$ is the size of a dust grain, 
$v_{\rm th}=\left(8/\pi\right)^{1/2}c_{\rm s}$ is the mean thermal velocity \citep{tak02}, $\Delta \mbf{v}$ is the relative velocity of dust to gas
and ${\cal M}$ is Mach number of the relative flow,
${\cal M}=|\Delta\mbf{v}|/c_{\rm s}$.
Equation (\ref{eq;epsori}) has the following asymptotic behavior
\citep{ahn76,paa07},
\begin{equation}
 \mbf{F}_{\rm D}=\left\{
		  \begin{array}{l}
		   -\frac{4\pi}{3}\rho_{\rm g}a^2v_{\rm
                    th}\Delta\mbox{\boldmath $v$} \ \ {\rm if}
                    \ |\Delta\mbf{v}|\ll c_{\rm s}\\
		   -\pi\rho_{\rm g}a^2|\Delta\mbf{v}|\Delta\mbf{v} \ \ {\rm
                    if} \ |\Delta\mbf{v}|\gg c_{\rm s}.
		  \end{array}
		 \right.
 \label{eq;epsasy}
\end{equation}
In the small $\Delta v$ limit, the drag force is from
stochastic collision by randomly moving gas particles, and thus,
the collision rate is proportional to the thermal velocity,  
which gives the dependence of $\mbf{F}_{\rm D}$ on $v_{\rm th}$.
On the other hand, in the large $\Delta v$ limit, the dominant factor
that determines the collision rate is the relative velocity between
dust and gas.
Therefore, the dependence on $v_{\rm th}$ in the drag force for the
small $\Delta v$ limit is replaced by the dependence on $\Delta v$ for
large $\Delta v$.
In this work, instead of using Equation (\ref{eq;epsori}),
we adopt a simplified formula that
interpolates between these limits \citep{kwo75,paa07},
\begin{equation}
 \mbf{F}_{\rm D}=-\frac{4\pi}{3}\rho_{\rm g}a^2v_{\rm
  th}f_{\rm d}|\Delta\mbf{v}|,
  \label{eq;epst}
\end{equation}
where
\begin{equation}
 f_{\rm d}=\sqrt{1+\frac{9\pi}{128}{\cal M}^2}. 
  \label{eq:fd}
\end{equation}
The stopping time for the gas drag is defined as $t_{\rm stop}\equiv m\Delta v/\left|\mbf{F}_{\rm D}\right|$,
where $m$ is the mass of a dust grain.
$t_{\rm stop}$ is a time scale for the decay of
the relative 
velocity between the dust and the gas. 
By normalizing with $\Omega_{\rm K}$, we define the nondimensional stopping time, 
$\tau_{\rm s}=t_{\rm stop}\Omega_{\rm k}$. 
In this paper we assume that the dust grains have spherical shape with homogeneous internal density, $\rho_{\rm m}$. 
Since the mass of a dust grain is 
$m=(4/3)\pi a^3\rho_{\rm m}$, then the nondimensional 
stopping time is rewritten as
\begin{equation}
 \tau_{\rm s}=\frac{\rho_{\rm m}a}{\rho_{\rm g}v_{\rm th}}f_{\rm
  d}^{-1}\Omega_{\rm k} .
  \label{eq;taus}
\end{equation}
For later discussion, we evaluate $\tau_{\rm s}$ at the midplane of the
MMSN. The relative velocity of dust to gas is
much smaller than the sound speed around the midplane.
Therefore, the nondimensional stopping time, $\tau_{\rm s,mid}$, at the midplane is

\begin{eqnarray}
  \tau_{\rm s,mid}& = & 
  1.8\times 10^{-7} 
  \left(\frac{a}{1\mu{\rm m}}\right)
  \left(\frac{\rho_{\rm g,mid}}{\rho_{\rm g,mid,1AU}}\right)^{-1} \nonumber \\
  & & \hspace{0.5cm}\left(\frac{c_{\rm S}}{c_{\rm s,1AU}}\right)^{-1}
  \left(\frac{\Omega_{\rm K}}{\Omega_{K,1{\rm AU}}}\right) \nonumber \\
  &=& 1.8\times 10^{-7}
  \left(\frac{a}{1\mu{\rm m}}\right)\left(\frac{r}{1\;{\rm AU}}\right)^{\frac{3}{2}}.
  \label{eq;ndstmid}
\end{eqnarray}
where we adopt $\rho_{\rm m}=2\;{\rm g\; cm^{-3}}$ in this work 
\citep[][]{you07}, and subscripts ``mid'' and ``1AU'' denote that the values 
are evaluated at the midplane and at $r=1$ AU, respectively. 
Here, we use, 
in addition to the relation for the Keplerian rotation, $\Omega_{\rm K}
=\Omega_{\rm K,1AU}(r/1{\rm AU})^{-3/2}$, the scalings for the density at the 
midplane and the sound velocity of the MMSN \citep{hay81,san00}: 
$\rho_{\rm g,mid} = \rho_{\rm g,mid,1AU}
\left(r/1{\rm AU}\right)^{-11/4}$ with $\rho_{\rm g,mid,1AU}=1.4\times 
10^{-9}$g cm$^{-3}$ and $c_{\rm s} = c_{\rm s,1AU}
\left(r/1{\rm AU}\right)^{-1/4}$ with $c_{\rm s,1AU}=0.99$ km s$^{-1}$. 
The nondimensional stopping time, 
$\tau_{\rm s}$, measures whether dust grains are tightly ($\tau_{\rm s}<1$) or weakly ($\tau_{\rm s}>1$) coupled to
the background gaseous flow.

The velocity of dust fluid is subject to the following momentum equation 
\citep{tak02}:
\begin{equation}
  \frac{dv_{\rm d}}{dt}=-\Omega_{\rm k}^2z-\frac{v_{\rm d}-v_{\rm g}}{t_{\rm stop}}+\frac{J}{\rho_{\rm d}t_{\rm stop}},
  \label{eq;deom}
\end{equation}
where $v_{\rm d}$
and $v_{\rm g}$
are the vertical velocity of dust and gas. 
The first term is the vertical component of the gravitational 
force by the central star, the second term
denotes the gas drag force on the dust from the mean gas flow, 
and the third term 
represents turbulent diffusion. 
$J$ is the
mass flux due to turbulent diffusion
and modeled as 
\begin{equation}
  J=-\frac{\rho_{\rm g}\nu}{Sc}\frac{\partial}{\partial z}\left(\frac{\rho_{\rm d}}{\rho_{\rm g}}\right)
  \label{eq:dmf}
\end{equation}
\citep{tak02}, where $\nu$ is effective kinetic viscosity of gas and
expressed as $\nu=\alpha H_0^2\Omega_{\rm k}/2$ with $\alpha$ parameter
\citep{ss73}.
Here, we adopt a constant $\alpha=0.01$ in the entire calculation region
for simplicity. The value of $\alpha$ controls the timescale of the upward
diffusion of grains that are located initially at $z=0$ to the wind region,
$z\gtrsim 3H_0$. However, the final profile of $\rho_{\rm d}$ in the
quasi-steady state does not depend on $\alpha$ provided that $\alpha$ is
finite, although smaller $\alpha$ requires longer time to achieve
the quasi-steady state.
The Schmidt number Sc is a ratio of gas and dust diffusion coefficients, $D_{\rm g}/D_{\rm d}$, which represents the strength of the coupling between the gas and the dust. For small grains, Sc approaches unity
so that
the grains have the same diffusivity as the gas, while it becomes
infinite for large grains. For intermediate grains, it is $Sc \approx
1+\tau_{\rm s}^2$ \citep[]{you07}.
In this paper, we use $Sc=1$, because
smaller grains with $\tau_{\rm s}<1$
are tightly coupled to the gas. 
Therefore, the steady state condition is achieved for the dust, 
and the dust velocity
reaches terminal velocity. Thus,
applying the time-steady 
condition ($\frac{d}{dt}=0$) to Equation (\ref{eq;deom}), 
we have
\begin{equation}
  v_{\rm d}=v_{\rm g}-\Omega_{k}\tau_{\rm s}z+\frac{J}{\rho_{\rm d}}.
  \label{eq;dtv}
\end{equation}
In addition, by substituting Equation (\ref{eq;dtv}) into the 
continuity equation of dust, 
$\frac{\partial \rho_{\rm d}}{\partial t}+\nabla\cdot\left(\rho_{\rm d} 
\mbf{v}_{\rm d}\right)=0$, we obtain

\begin{equation}
  \frac{\partial\rho_{\rm d}}{\partial t}+\frac{\partial}{\partial z}\left(\rho_{\rm d}V_{\rm d}+J\right)=0, 
  \label{eq;eoc} 
\end{equation}
where we separate the dust velocity $v_{\rm d}$ into the mean flow,
\begin{equation}
  V_{\rm d}=v_{\rm g}-\Omega_{\rm k}\tau_{\rm s}z,
  \label{eq;terv}
\end{equation}
and the diffusion component, $J/\rho_{\rm d}$.

By solving Equation (\ref{eq;eoc}), we follow the time evolution 
of the vertical distribution of the dust fluid.

\subsection{Gas Component}
\label{sec:gas}

We adopt the result of the three-dimensional (3D) MHD simulation in a local 
shearing box by \citet{SI09} for the fixed background gas component. 
In order to incorporate it into our vertical 1D calculation, the time- and 
horizontal plane-average is taken for $v_{\rm g}$.

Since our computation region, $|z|\le10H_0$, is larger than the region, 
$|z|\le\pm 4H_0$, of that 3D simulation, we
have to
extrapolate $v_{\rm g}$ and 
$\rho_{\rm g}$ to $|z|>4H_0$. To do so, we derive an analytic expression 
that fits to the $v_{\rm g}$ profile of the 3D simulation:

\begin{equation}
  v_{\rm g}=50c_{\rm s}\left(\tanh\left(\frac{z-5.6}{0.62}\right)+1\right).
  \label{eq;fit}
\end{equation}
By this fitting formula, we bound the flow velocity $\le 100 c_{\rm s}$ in the 
high-altitude region in order to avoid unrealistically high-speed flows.

\begin{figure}[!h]
  \centering
  \includegraphics[width=7.5cm,bb=85 50 300 300,clip]{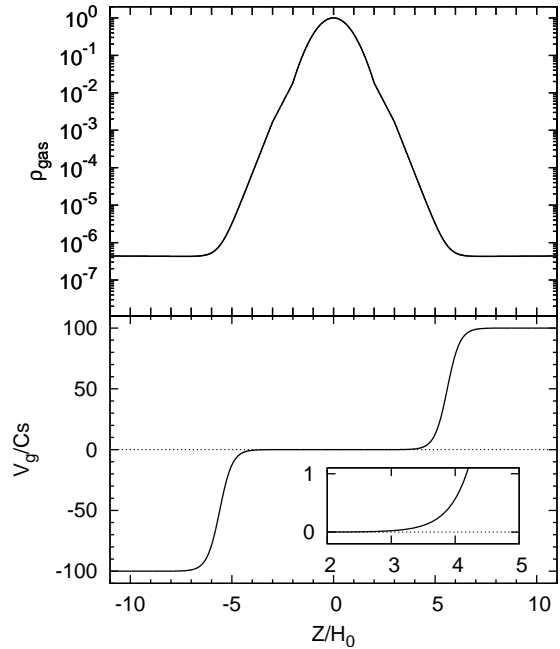}
  \caption{
    Vertical distribution of density ($\rho_{\rm g}$; {\it top}) 
    and velocity normalized by the sound speed ($v_{\rm g}/c_{\rm s}$; 
    {\it bottom}) of the background gas component.
  }
  \label{fig;gdv}
\end{figure}

As for the density distribution, we adopt the following procedure. 
In $|z|<2H_0$, we use the hydrostatic density structure, $\rho_{\rm g}= 
\rho_{\rm g,mid} \exp (-z^2/H_0^2)$, which
well explains the 3D MHD simulation. 
In $|z| > 3H_0$, the 3D simulation shows that the mass flux $\rho_{\rm g} 
v_{\rm g}=4\times 10^{-5}\rho_{\rm g,mid}c_{\rm s}$ is constant with $z$. 
Then, we derive $\rho_{\rm g}$ from 
Equation (\ref{eq;fit}) to give the constant $\rho_{\rm g} v_{\rm g}$.
We linearly connect these two regions between 
$z=\pm2H_0$ and $z=\pm3H_0$. 
Reflecting that $v_{\rm g}$ approaches $100 c_{\rm s}$ in the upper 
regions, $|z|>6H_0$, the density is also bounded by $\rho_{\rm g}\ge 4\times 
10^{-7}\rho_{\rm g,mid}$ in these regions.
Figure \ref{fig;gdv} presents the derived density and velocity 
of the gas component.

\section{Results}
\label{sec:res}
\subsection{Time evolution}

\begin{figure}[!h]
 \centering
 \includegraphics[width=8cm,clip]{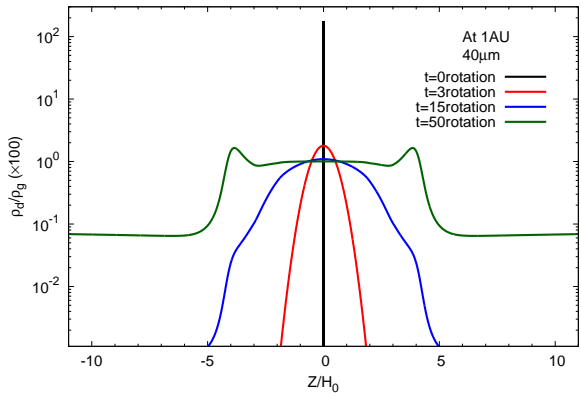}
 \caption{
 Time evolution of the dust-to-gas ratio, $\rho_{\rm d}
 /\rho_{\rm g}$ for the dust
 grains of uniform size, 
 $a=40\mu$m, at 1 AU.
 (Stopping time, $\tau_{\rm s,mid}=7.2\times10^{-6}$)
 Note that the value of $\rho_{\rm d}/\rho_{\rm g}$ is multiplied by 100.
 The red, blue, and green lines 
 correspond to rotation time $=3$, 15, and 50, respectively.
 After t$>$50 rotations, the system 
 is in the quasi-steady state.
 }
 \label{fig;tme}
\end{figure}

Figure \ref{fig;tme} demonstrates the time evolution of dust density, 
$\rho_{\rm d}$, for a single grain size, $a=40\mu$m, at $r=1$ AU of the MMSN, 
or $\tau_{\rm s,mid}=7.2\times 10^{-6}$
The figure clearly exhibits the dust
grains initially located at the midplane are
stirred upward with time. 
After approximately three rotations (red line), the dust grains are 
diffused by the turbulence up to $z=\pm2H_0$ 
where the disk winds launch.
Around
$t\sim 15$
rotations (blue line), 
the local peaks of the dust-to-gas ratio gradually form at 
$z\approx \pm 4 H_0$. 
The locations of the peaks approximately correspond to the sonic
points of the background gaseous disk wind where the drag force switches
from the small $\Delta v$ regime to the large $\Delta v$ regime
(Equation \ref{eq;epsasy}; see Section \ref{sec:depgs}
for more details).
The location of the $\rho_{\rm d}/\rho_{\rm g}$ peaks depends 
sensitively on the size of dust grains. Although in this paper we assume the 
spherical shape and the uniform solid density for grains, it also depends on 
the properties of dust in realistic situations.
The dust-to-gas ratio further increases with time, and it 
saturates after $t=50$ rotations (green line) to be in the quasi-steady 
state. The region inside the $\rho_{\rm d}/\rho_{\rm g}$ peaks, the dust-to-gas 
ratio approaches the volumetric average value $=0.01$ because of 
the well coupling between the gas and the dust.
We continue the simulation further up to 200 rotations to study the 
quasi-steady-state behavior, which we discuss
in the subsequent sections.

\subsection{　Dependence on Grain Size}
\label{sec:depgs}

\begin{figure*}[t]
  \centering
  \begin{tabular}{cc}
    \includegraphics[bb=80 50 300 300,clip]{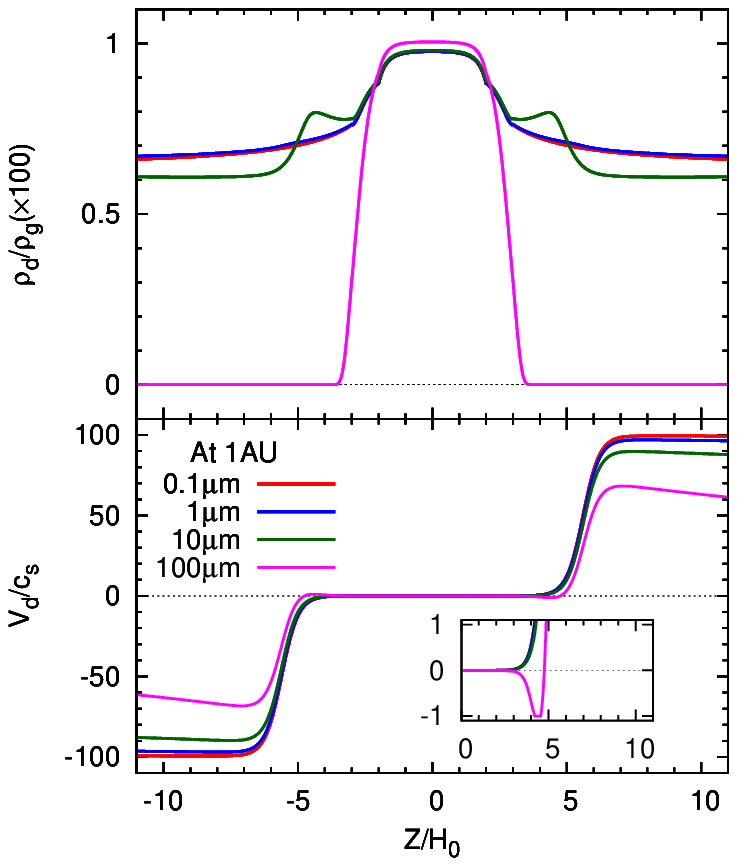}
    \includegraphics[bb=82 50 300 300,clip]{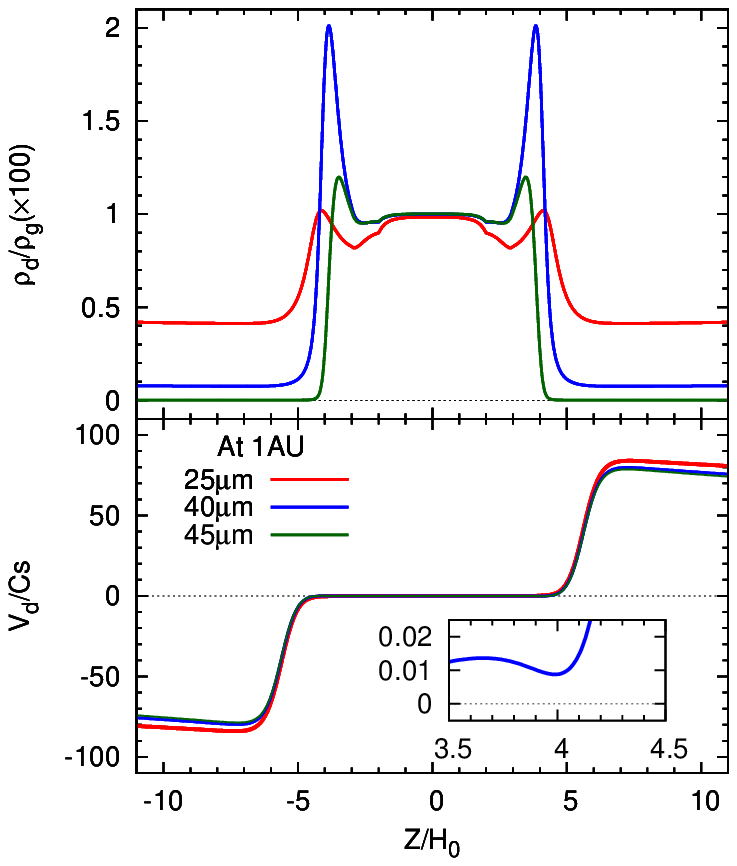}
  \end{tabular}
    \caption{Time-averaged vertical structure of the dust
      distribution with various $\tau_{\rm s,mid}$ during $t=100-200$ rotations. 
      The top panels show the dust-to-gas ratio, and the bottom panels 
      show the mean velocity (Equation \ref{eq;eoc}). The inset in the bottom panels 
      zooms in the region near $V_{\rm d}=0$.
      The left panels present the results with $\tau_{\rm s,mid}=1.8\times10^{-8}$ 
      (red), $1.8\times10^{-7}$ (blue), $1.8\times10^{-6}$ (green), and
      $1.8\times10^{-5}$(pink), which 
      correspond to $a=$0.1, 1, 10, and 100 $\mu$m at 1 AU of the MMSN. 
      The right panels present the cases with
      $\tau_{\rm s,mid}=4.5\times10^{-6},7.2\times10^{-6},8.1\times10^{-6}$,
      which correspond to
      $a=$25, 40, 45 $\mu{\rm m}$ and show the $\rho_{\rm d}/\rho_{\rm g}$ peaks in
      $|z|\sim4 H_0$. 
  }
  \label{fig;dndst}
\end{figure*}

We examine the dependence of the vertical dust distribution on 
grain size, $a$, or stopping time, $\tau_{\rm s, mid}$.
Figure \ref{fig;dndst} presents the time-averaged steady-state vertical 
profile of the dust-to-gas ratio (top panels) and the mean flow velocity of 
the dust, $V_{\rm d}$, (Equation (\ref{eq;terv}); bottom panels) for different 
stopping times. The left panels show the cases with 
$\tau_{\rm s,mid}=1.8\times10^{-8}$, $1.8\times10^{-7}$, $1.8\times10^{-6}$, and $1.8\times10^{-5}$, which respectively correspond 
to the grain size of $a=$0.1, 1, 10, and 100 $\mu$m at $r=$1 AU of the MMSN.

In the region around the midplane, $|z|\lesssim 2H_0$, the 
dust-to-gas ratio is consistent with the volumetric average value, 
$\rho_{\rm d}/\rho_{\rm g}=0.01$, which is independent from $\tau_{\rm s,mid}$, 
because the dust is strongly coupled with the gas. However, the value of 
$\rho_{\rm d}/\rho_{\rm g}$ depends largely on $\tau_{\rm s,mid}$ in the 
high-latitude regions, since the coupling is weaken (the local 
$\tau_{\rm s}$ increases) owing to the decrease of the gas density with 
elevating height. The dust with the small grain size, 
$\tau_{\rm s,mid}<10^{-6}$, is well coupled to the gas in the entire simulation 
domain, and is blown upward with the vertical outflows of the gas. 
On the other hand, the dust with the large grain size, 
$\tau_{\rm s,mid} >10^{-5}$, remains in the disk.
The velocity of the mean dust flow, $V_{\rm d}$, in the wind region 
is faster for a smaller grain size (smaller $\tau_{\rm s}$) owing to the well 
coupling to the gas. In the case with the largest grain size, 
$\tau_{\rm s,mid}=1.8\times10^{-5}$, $V_{\rm d}$ is directed to the midplane, which is 
however compensated for by the diffusion, $J/\rho_{\rm d}$ in 
Equation (\ref{eq;deom}). Small grains escape from the simulation domain, 
which leads to the loss of the total mass of the dust. However, the mass loss 
is almost negligible 
during the simulation time, $t\le 200$ rotations, 
and then, it is reasonable to treat the system is in the quasi-steady state 
even for the well-coupled small grains.

Among the four cases, the dust-to-gas ratio of the case with 
$\tau_{\rm s,mid}=1.8\times10^{-6}$ shows peculiar behavior; $\rho_{\rm d}/\rho_{\rm g}$ 
increases with an altitude in the high-altitude layers,
$2.5H_0\lesssim |z|\lesssim 4.5H_0$.  
On the other hand, the dust remains in the disk and $\rho_{\rm d}=0$ 
in the high-altitude layers of the case with $\tau_{\rm s,mid}=1.8\times10^{-5}$.  
In the right panels of Figure \ref{fig;dndst}, we focus on grain sizes 
with $\tau_{\rm s,mid}=4.5\times10^{-6}, 7.2\times10^{-6}, 8.1\times10^{-6}$, 
which lie between these two cases.
One can see the peaks of $\rho_{\rm d}/\rho_{\rm g}$
near $z\approx \pm 4 H_0$ in these cases
(right top panel), which indicates that the dust
grains float up to the location around the peaks and stagnate there.
The emergence of the peaks can be explained for dust grains of
$\tau_{\rm s}=7.2\times10^{-6}$ in the following.

The inset of the bottom right panel of Figure \ref{fig;dndst}
zooms in the dust velocity, $V_{\rm d}$, near the location of the peak
in the positive $z$ side.  $V_{\rm d}$ slightly decreases from
$z\approx 3.7H_0$ to $\approx 4H_0$, and  
shows a local minimum at $z\approx 4H_0$, which almost coincides with the sonic
point at $z=4.2H_0$ of the gaseous disk wind.
A quasi-steady state is achieved in the simulation, and thus,
$\rho_{\rm d}V_{\rm d}\approx$ const. Therefore, the local minimum of
$V_{\rm d}$ leads to a local maximum of $\rho_{\rm d}$; dust grains
stagnate around this location.

The non-monotonic behavior of $V_{\rm d}$ can be explained by the
variation of the upward drag force against the downward gravity. 
Below the location of the $\rho_{\rm d}/\rho_{\rm g}$ peak at
$|z|\approx 4H_0$, the velocity of the gaseous disk wind is still
subsonic, $v_{\rm g}<c_{\rm s}$. The relative velocity, $\Delta v$,
of the dust to the gas is also $< c_{\rm s}$, and the drag force is in the
small $\Delta v$ regime in Equation (\ref{eq;epsasy}). Therefore,
$f_{\rm d}\approx 1$ (Equation \ref{eq:fd}), and $\tau_{\rm s}$
increases with height because of the decrease of $\rho_{\rm g}$:
\begin{equation}
  \hspace{-0.6cm} \tau_{\rm s}(z) \approx
  \tau_{\rm s,mid}\frac{\rho_{\rm g,mid}}{\rho_{\rm g}(z)}
  =0.18\left(\frac{\tau_{\rm s,mid}}{7.2\times 10^{-6}}\right)
  \left(\frac{\rho_{\rm g}(z)/\rho_{\rm g,mid}}{4\times
    10^{-5}}\right)^{-1},
  \label{eq;tausubsonic}
\end{equation}
where we normalize with the gas density,
$\rho_{\rm g}=4\times 10^{-5}\rho_{\rm g,mid}$ at the location,
$|z|\sim4H_0$, of the local minimum $V_{\rm d}$.
In $|z|\lesssim 4H_0$, the coupling between the gas and
the dust is lessened with an increasing height. 
Therefore, the downward gravity
gradually dominates the upward gas drag force,
which causes the deceleration of $V_{\rm d}$ in $3.7H_0 < |z| < 4H_0$.

However, $V_{\rm d}$ rapidly increases beyond $|z|>4H_0$.
The relative velocity, $\Delta v$, of the dust to the gas exceeds
$c_{\rm s}$ beyond the sonic point of $v_{\rm g}$ located at $|z|=4.2H_0$.
Then, the drag force switches to the large $\Delta v$ regime in Equation
(\ref{eq;epsasy}). In the large $\Delta v \gg c_{\rm s}$ limit,
$f_{\rm d} \propto {\cal M}$ (Equation \ref{eq:fd}), which gives
$\tau_{\rm s}\propto (\rho_{\rm g} v_{\rm th} f_{\rm d})^{-1} \propto
(\rho_{\rm g}\Delta v)^{-1}\sim (\rho_{\rm g} v_{\rm g})^{-1}=$ const.
(Equation \ref{eq;taus}), where we here assume
$\Delta v=|v_{\rm g}-v_{\rm d}|\sim v_{\rm g}$.
This shows that $\tau_{\rm s}$, and therefore $t_{\rm stop}$,
does not increase (the coupling between gas and dust is not weaken),
with an increasing altitude in the large $\Delta v$ regime.
As a result, the upward gas drag force increases with the increasing
$v_{\rm g}$ (the 2nd term of Equation \ref{eq;deom}) and dominates
the downward gravity (the 1st term of Equation \ref{eq;deom}, which
gives the rapid acceleration of $V_{\rm d}$ in $|z|>4H_0$.

As explained above, the local peaks of the
$\rho_{\rm d}/\rho_{\rm g}$ ratio emerges when dust grains satisfy
the condition that they are loosely coupled ($\tau_{\rm s}\sim 0.1-0.2$)
to the background gas near the sonic point of the disk wind.
By calculating many cases with different stopping times, the range of
$\tau_{\rm s,mid}$ that gives $\rho_{\rm d}/\rho_{\rm g}$ peaks is 
\begin{equation}
 \tau_{\rm s,mid}\approx(4.5-8.1)\times10^{-6}.
  \label{eq;taupeak}
\end{equation}
Smaller grains escape from the simulation domain, because
they are well coupled to the gas flow
even near the sonic point. 
Larger grains remain in the subsonic region of the disk wind
because they cannot be lifted up beyond the sonic point where the effective
drag force in the large $\Delta v$ regime could act.
Therefore, dust grains outside the range of Equation (\ref{eq;taupeak}) do not
show local $\rho_{\rm d}/\rho_{\rm g}$ peaks.

\subsection{Radial dependence}

\begin{figure}[!h]
  \centering
  \includegraphics[width=7cm,clip]{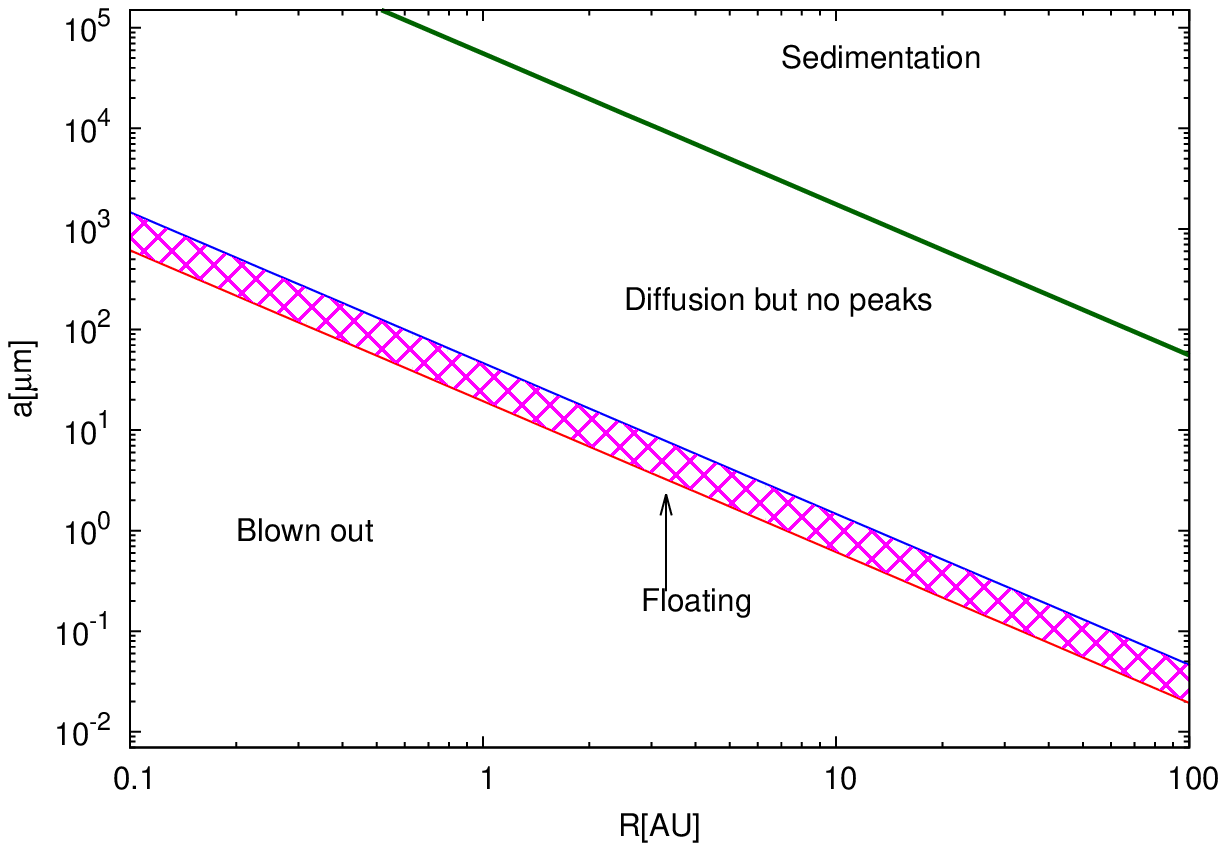}
    \includegraphics[height=7cm,width=8cm,clip]{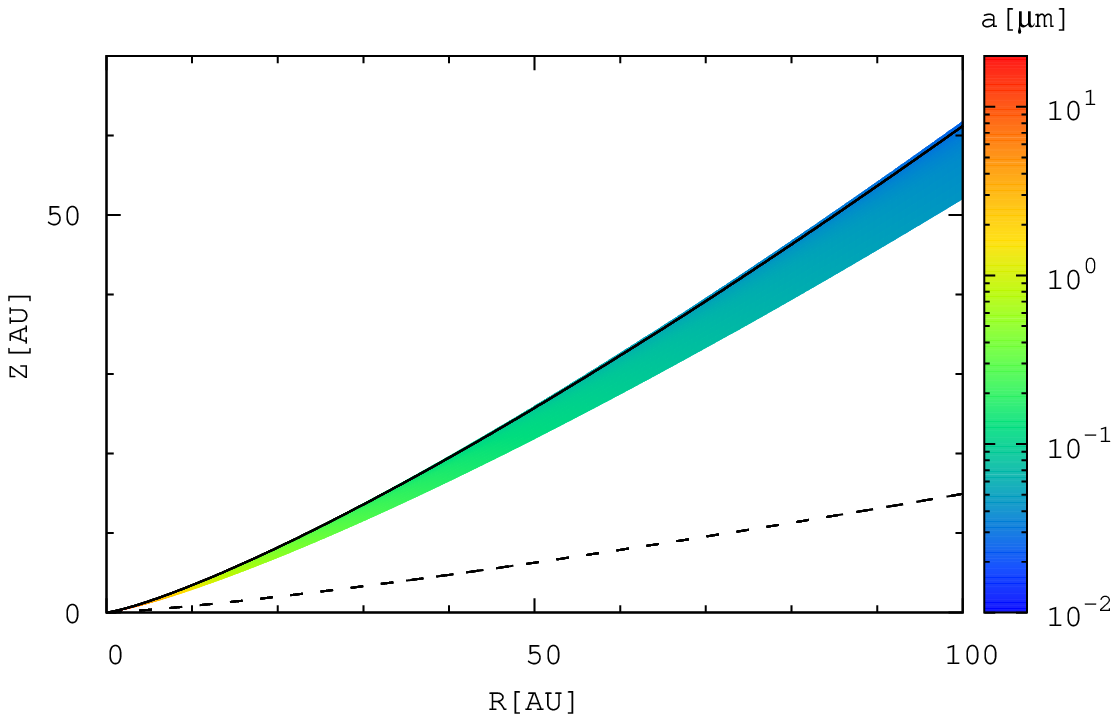}
  \caption{Properties of the
    dust grain dynamics in the MMSN with the disk wind. 
 {\it Top: }Behavior of the dust with different grain size, $a$, on radius 
 from the central star. The pink hatched region corresponds to the dust 
 particles that float in the high-altitude layer 
 and show a sharp concentration of $\rho_{\rm d}/\rho_{\rm g}$
 around $|z|\approx 4H_0$. 
 The dust particles below the pink hatched region are blown away with the 
 gas outflows from the simulation domain.
 Those between the pink hatch and the green line are distributed in the disk without 
 $\rho_{\rm d}/\rho_{\rm g}$ peaks but are stirred up to $|z|>1 H_o$. 
 Those above the green solid line are settled within $1 H_0$ around the 
 midplane. {\it Bottom:} Floating dust grains in an $r-z$ diagram. Color contour 
 indicates the grain size in the pink hatched region 
 in the top panel.
 The black solid line
 indicates the sonic point of the gaseous disk
 wind, and the dotted line
 denotes one scale height, $z=1 H_0$.
 }
  \label{fig;rdep}
\end{figure}

The result based on the local vertical box (Figure \ref{fig;dndst}) 
has shown that the behavior of the dust is determined by the nondimensional 
stopping time, $\tau_{\rm s, mid}$, at the midplane. 
Therefore, once we specify the 
radial dependence, we can determine the dust properties in an $r-z$ plane. 
For illustrative purpose,
we adopt the MMSN (see Equation \ref{eq;ndstmid} and below),
and show the basic character of the vertical motions of dust grains
as a function of the size $a$ and the distance from the central star
in Figure \ref{fig;rdep}.
The pink hatched region in the top panel 
indicates the radial dependence of
the size, $a$, of the dust grains
that floats in the high-altitude layers 
with the $\rho_{\rm d}/\rho_{\rm g}$ peaks. Reflecting the $r^{3/2}$ dependence 
of $\tau_{\rm s,mid}$ (Equation \ref{eq;ndstmid}), the size of 
the floating dust grains is smaller for larger distance,
\begin{equation}
a_{\rm float} \approx (25-45)\; \mu{\rm m}
\left(\frac{r}{\rm 1\; AU}\right)^{-3/2}.
\label{eq:raddep}
\end{equation}
The dust grains above the pink hatched 
region basically remain in the disk, while those below the pink hatch are blown 
upward with the vertical gas flow.

The bottom panel of Figure \ref{fig;rdep} illustrates the pink hatched region 
in the top panel in an $r-z$ plane.
This panel clearly shows that
dust grains are floating in a narrow range near or slightly below
the sonic point of the background gaseous disk wind.
The size of the floating grains decreases for larger $r$. 
Smaller-size grains remain only in an outer region of the disk, while 
larger-size grains are distributed in a broader region. For instance, grains with 
$a=0.1\mu$m remain in $r\gtrsim 50$ AU, while grains with $a=100\mu$m 
remain in $r\gtrsim 0.5$ AU. If we consider the time-evolution of a 
protoplanetary disk, the dispersal of the dust begins with small-size grains 

If we focus on dust grains with a certain size, say $a=1\mu$m, they are blown 
away by the vertical gas outflow in the inner region, $r\lesssim 10$ AU, 
and remain in the disk in the outer region, $r\gtrsim 10$ AU. 
Therefore,
we expect that
the evacuation of dust grains is expected to occur in an inside-out manner, 
which is qualitatively similar to the evolution of the gas component
driven by
the disk wind \citep{SMI10}.

\section{DISCUSSION}
  \subsection{Effect of Reduced Mass Flux of the Disk Wind}

  In this paper
  we use a simple model that adopts the time-
  and horizontal-plane-averaged gas density and velocity derived from the local 
  3D shearing box simulation solving ideal MHD equations \citep{SI09} for the 
  background flow.

\begin{figure}[!h]
   \centering
   \includegraphics[width=7.5cm,clip]{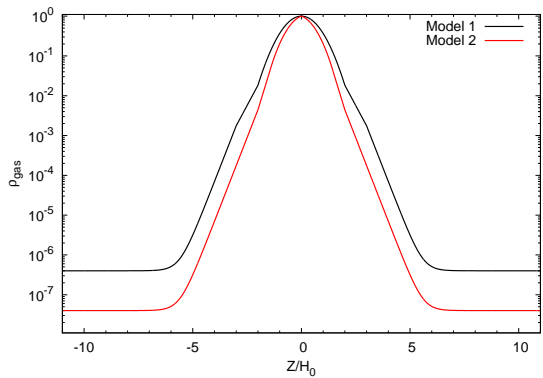}
   \caption{
   Comparison of the gas densities of Models 1 and 2.
   }
   \label{fig;mdldif_gdv}
  \end{figure}

  However, there is uncertainty in the mass flux of disk
  winds calculated by shearing box simulations.
  In particular, the outflow mass flux
  shows an decreasing trend on an increasing vertical box size
  \citep{SMI10,fro13}, which implies that the mass flux of the disk wind
  we have used so far might overestimates an actual value.
  Therefore, we here check how the distribution of the dust is
  modified when the outflow mass flux of the gas is smaller.
  In comparison to the original setup, which we call
  ``Model 1'' hereafter, we adopt 10 times smaller vertical mass flux of
  the gas component, $\rho_{\rm g} v_{\rm g}$, in ``Model 2'',
  by reducing $\rho_{\rm g}$ by 1/10 in the
  wind region (Figure \ref{fig;mdldif_gdv})
  but leaving $v_{\rm g}$ unchanged.

\begin{figure}[!h]
   \centering
   \includegraphics[width=7.5cm,bb=85 50 300 300,clip]{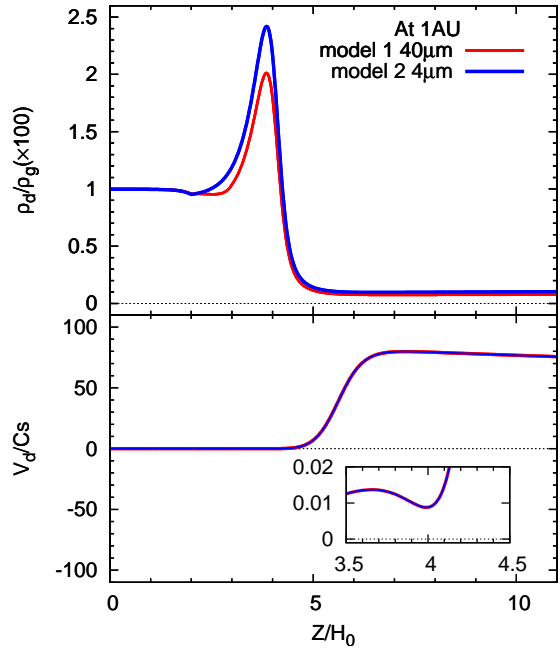}
   \caption{The same as the right panels of Figure
   \ref{fig;dndst} but for
   the comparison between Models 1 and 2.
   }
   \label{fig;mdldif_dstpeak}
  \end{figure}

  Figure \ref{fig;mdldif_dstpeak} presents the time-averaged
  steady-state vertical profile of the dust-to-gas ratio (top panel) and
  the mean flow velocity of the dust, $V_{\rm d}$, (bottom panel)
  of Model 2 in comparison to those of Model 1.
  The gas density of Model 2 is 1/10 of that of Model 1 
  in the supersonic region. Since $\tau_{\rm s}\propto 1/\rho_{\rm g}$
  (Equation \ref{eq;taus}), it is expected that in Model 2 dust grains
  with smaller $\tau_{\rm s,mid}$ at the midplane will give similar
  results to Model 1 in the supersonic region. Therefore, we adopt
  $\tau_{\rm s,mid}=7.2\times 10^{-7}$ in Model 2, which is 1/10 of
  $\tau_{\rm s,mid}(=7.2\times 10^{-6})$ of Model 1, whereas 
  $\tau_{\rm s,mid}=7.2\times 10^{-6}$ and $7.2\times 10^{-7}$ correspond to
  the grain size of $a=40\mu$m and $a=4\mu$m at $r=1$ AU of the MMSN,
  respectively.

  Figure \ref{fig;mdldif_dstpeak} shows that the velocity profiles
  of these two cases are almost the same, and so are the locations of the
  peaks. This indicates that the dust floating by the stagnation of upflowing
  grains is a robust feature
  in a qualitative sense even though the mass flux
  of the gaseous disk wind might be reduced, whereas the size of floating dust grains
  is smaller for smaller mass flux of the disk wind.
  Model 2 shows larger $\rho_{\rm d}/\rho_{\rm g}$ in the subsonic region,
  $|z|<4H_0$, because the gas density of Model 2 there is not as low as 1/10
  of that of Model 1. Thus, $\tau_{\rm s}$ in the subsonic region of
  Model 2 is smaller than $\tau_{\rm s}$ of Model 1,
  and a larger amount of dust grains are lifted up by
  more efficient coupling to the gas.

  \subsection{Limitations}

  In the previous subsection, we have discussed the uncertainty of
  the wind mass flux in the local shearing box approximation. In addition to
  this, the configuration of large-scale magnetic field is another effect
  which cannot be taken into account in the local treatment.
  Gas flows along large-scale
  magnetic field lines, and the mass conservation
  should follow $\rho_{\rm g}v_{\rm g}/B_{\rm p}=$ const. in realistic situations,
  where $B_{\rm p}$ is poloidal magnetic field.
  Thus, the mass flux, $\rho_{\rm g} v_{\rm g}$, would probably
  decreases as the magnetic flux tube expands with altitude, and hence
  $B_{\rm p}$ decreases.
  As a result, the
  location and detailed shape of $\rho_{\rm d}/\rho_{\rm g}$ peaks might be
  modified in such situations. However, the physical mechanism that forms
  a $\rho_{\rm d}/\rho_{\rm g}$ peak near the sonic point of vertical upflows
  (Section \ref{sec:res}) is robust in a qualitative sense.

  In this paper, we have assumed the steady-state condition for the background
  gaseous flows. In reality, however, the disk wind is intermittent with
  quasi-periodicity of 5-10 rotations \citep{SI09,SI14}.
  We need to take into account
  effects of time-dependent vertical upflows on dust grains in more detailed
  studies \citep{tur10}.

  The strong coupling approximation adopted in this paper (Equation
  \ref{eq;dtv}) is basically acceptable for floating dust grains because
  $\tau_{\rm s}(\approx 0.2)$ is kept less than unity near and above
  the location where they are floating (Section \ref{sec:res}).
  However, if time-dependent vertical upflows were taken into account,
  grains with $\tau_{\rm s}\gtrsim 1$ might appear locally in a
  transient manner. For those grains, it is necessary to solve the
  time-dependent equation of motion (Equation \ref{eq;deom}). 

  We treat dust grains as test particles, because we start the
  simulations from the total dust-to-gas ratio $=0.01$, and the condition,
  $\rho_{\rm d}/\rho_{\rm g}<1$, is satisfied in the throughout the simulations.
  However, when we consider the depletion of the gas component at later
  evolutionary stages, 
  $\rho_{\rm d}/\rho_{\rm g}$
  might exceeds unity especially near the peak location.
  In order to treat the long-time evolution of protoplanetary disks,
  the backreaction from the dust on the gas \citep[e.g.,][]{jk05}

  We use the result of the ideal MHD simulation, namely we do not consider the 
  effect of an MRI-inactive region \citep[``dead zone''; e.g.,][]{gam96}.
  If the effects of non-ideal MHD are taken into account, the
  turbulence strength, $\alpha$, would be smaller than the adopted value,
  $\alpha=0.01$, and thus it takes time for grains at the midplane
  at the beginning of the calculations to diffuse upward (see Section
  \ref{sec:dust}).
  After the quasi-steady state is achieved,
  however, a possible existence of dead zone would not change our result on floating
  dust grains at least in a
  qualitative 
  sense, because the disk wind is driven from the MRI-active region near the 
  surface where the sufficient ionization is attained.
  For instance, \citet{SMI10} demonstrated that the mass flux of the disk 
  wind is weaken by half for a typical case with a dead zone.
  Recent simulations that include various non-ideal MHD effects
  also show that magnetically driven winds are launched in spite of the strong
  magnetic diffusion \citep[e.g.,][]{bs13b,Gre15}.
  Although the size of floating grains might be
  smaller than that predicted from our results based on the ideal MHD
  simulations, floating dust grains would be a robust feature even
  if non-ideal MHD effects were considered.

  When we consider dust grains 
  with $a=0.1\mu$m at $r=1$ AU, the dispersal timescale is $\approx 7000$ yrs, 
  which is much shorter than the typical lifetime of protoplanetary disks, 
  $\approx 10^{6-7}$ yrs \citep{hai01}.
  The main reason of this discrepancy is that the effect of the radial 
  accretion cannot be taken into account in the local box. In 
  realistic situations, dust grains would be supplied from the outer region 
  and the actual dispersal time would be much longer. \citet{SMI10}
  have done such
  similar calculations for the gas component and showed that the dispersal of 
  the disk gradually takes place with timescale of $\sim 10^6$ yrs even though 
  the dispersal time at $r=1$ AU is $\approx 4000$ yrs in the local treatment. 
  The similar argument is applicable to the dust component, and we expect that 
  the dispersal timescale of small dust grains
  will turn out to be
  a reasonable value in a global model.

   Since we have restricted our calculations to those in the local
    vertical box in this paper, local concentrations of floating dust grains
    are formed by the combination of the vertical component of the gravity
    and the drag force. In realistic situations, however, the direction of
    the flow, the gravity, and the drag force are not generally coaligned
    each other.
   In such circumstances, upflowing grains could play a role in
  large-scale circulation of the dust in protoplanetary disks. 
  The disk wind of the gas component is probably accelerated by magneto-centrifugal
  force to the radially outward direction \citep[e.g.,][]{bp82}.
  Dust grains that are well coupled to the gas are uplifted and transported
  to larger $r$ along large-scale magnetic field by such gaseous
  upward and outward flows.
  Eventually, the ambient gas density decreases as the magnetic
  flux tube expands with altitude, which makes the coupling to the gas 
  weaken
  to give $\tau_{\rm s} >1$. 
  Those grains will settle down to the midplane by the vertical
  component of the central-star gravity without sufficient gas drag force
  in such a distant region.
  {
  Outflows are expected to blown out intermittently \citep{SI09}, and
  hence, dust grains are lifted up and settled down  in a time-dependent
  manner in realistic situations, rather than the formation of
  time-steady local peaks discussed in the previous sections. The dust
  grains settled down at larger r further accrete inward, being coupled
  to the accreting gas. Dust grains are also supposed to grow through
  this large-scale circulation \citep{shu96}.
  }

\subsection{Observational Implications}

Finally we discuss observational implications of floating dust
in protoplanetary disks.
\citet{McJ14} estimated the column densities of neutral hydrogen
around many T-Tauri Stars and Herbig Ae/Be stars by
the circumstellar adsorptions of the Lyman-$\alpha$
emissions from the central stars.
The column densities derived from Lyman-$\alpha$ absorption
are smaller than the values derived from the dust extinction measurements.
A possibility of a high dust-to-gas ratio
in the upper-atmosphere of the protoplanetary disks
was proposed as a possible solution to the discrepancies.
\citet{ell14} observed outflows/winds from a Herbig Ae star.
The object shows variability where optical fading events
are correlated with near-infrared brightening,
which implies that a sizable amount of the dust grains
are contained in the variable outflows/winds.
The mechanism studied in this paper may provide an explanation
for existence of those dusty materials
in upper atmospheres of protoplanetary disks.

Also, 
the scattered light image in circumstellar
environment of RY Tau by near-infrared coronagraphic imaging polarimetry
indicated that the observed
polarized intensity distribution shows a butterfly-like distribution as shown
by the observation at millimeter wavelengths \citep{tkm13}. 
They performed comparison between the observed polarized intensity
distribution and models
consisting of gaseous disk and dust.
The result suggests the scattered light in the
near-infrared is associated with an optically thin and geometrically thick
layer at an upper layer.
This phenomenon could be explained by
the dust grains lifted up by the disk wind.   
In our future studies, we plan to create pseudo observational images
\citep[e.g.,][]{tkm14} based on a theoretical model for floating dust grains.

\section{SUMMARY}

We have studied the dynamics of dust grains 
in the protoplanetary disk winds driven by the MRI-triggered 
turbulence. Our calculation in the 1D vertical box has shown that the 
intermediate-size dust grains with $\tau_{\rm s,mid}\sim 7.2\times 10^{-6}$ 
float at about 4 scale heights from the midplane.
This location corresponds to the sonic point of the background
gaseous disk wind, where the drag force from the gas to the dust changes
from the small relative velocity regime to the large relative velocity regime.
The size of the floating dust grains depends on the distance 
from the central star; if we take the MMSN for example, we have 
size of floating dust,
$a_{\rm float}\propto r^{-3/2}$; smaller grains float at larger $r$. 
The dust grains with the smaller size than that of 
the floating ones
are well coupled to the bulk gas flow and continue to stream
out of the simulation domain, while the larger dust grains remain in the disk; 
this mechanism selectively disperses small grains in an inside-out 
manner and leaves large grains in protoplanetary disks.

One of the interesting findings in this paper is
the characteristic size of the dust grains floating
and stagnating at an upper atmosphere of the protoplanetary disk.  
The size range of those grains is a function of the distance
from the central star, i.e., $\rho_{\rm g}(r), c_{\rm s}(r)$
and appears to be relatively narrow as shown in the top panel of Figure 4. 
Future observations of the emission
and/or absorption by those floating dust grains may provide information
about the typical size of those grains; i.e.,
{\it the detection of a dust floating layer may immediately enable us
to determine the size of those particles}.  
Combining with the possible observations of the disk temperature
, i.e., $c_{\rm s}(r)$ and density,
we will be able to constrain the mass flux of bulk flow, $\rho_{\rm g}v_{\rm g}$,
which is difficult to observe directly. 
This may provide important observational diagnostics
on the dispersal timescale of individual protoplanetary disks.

\acknowledgments
We thank an anonymous referee for many valuable comments.
We also thank Hiroshi Kobayashi and Neal J. Turner for fruitful discussion.
S.I. are supported by Grant-in-Aid for Scientific Research (23244027,
23103005).
The work is supported by the Astrobiology Center Project of National
Institutes of Natural Sciences (NINS) (Grant Number AB271020).

\bibliography{ms_asph201602}

\end{document}